\documentclass[english,twoside,a4paper,11pt]{article}

\usepackage[utf8]{inputenc}
\usepackage[T1]{fontenc}
\usepackage{amsmath}
\usepackage{amsfonts}
\usepackage{graphicx}
\usepackage{subcaption}
\usepackage{a4wide}
\usepackage{amssymb}
\usepackage{fancyhdr}
\usepackage{mathrsfs}
\usepackage{amssymb}
\usepackage[toc,page]{appendix}
\usepackage{xcolor}
\usepackage{float}
\usepackage{cite}
\usepackage{hyperref}
\usepackage{fontawesome}
\usepackage[dvipsnames]{xcolor}
\usepackage{authblk}
\usepackage{graphicx}
\usepackage[skip=0.333\baselineskip]{caption}
\usepackage{subcaption}
\usepackage{url}
\usepackage{comment}

\def\ba{\begin{eqnarray}}
\def\ea{\end{eqnarray}}

\def\be{\begin{equation}}
\def\ee{\end{equation}}

\begin{document}

\title{\centering{Gray-body factors: Method matters II,\\ WKB vs GrayHawk}}

\author[1,2]{Marco Calz\'a}
\affil[1]{Dipartimento di Fisica, Universit\`a di Trento,
Via Sommarive 14, 38123 Povo (TN), Italy}
\affil[2]{Univ Coimbra, Faculdade de Ci\^encias e Tecnologia,
da Universidade de Coimbra and CFisUC,
Rua Larga, 3004-516 Coimbra, Portugal}

\author[3,4]{Andrea Spina}
\affil[3]{Department of Physics and Astronomy, Universit\`a di Catania, Via S. Sofia 64, 95123 Catania, Italy}
\affil[4]{INFN, Sezione di Catania,  Via S. Sofia 64, 95123, Catania, Italy.}

\date{}

\maketitle
\begin{abstract}
We present a systematic comparison between two independent approaches for the computation of gray-body factors. A fully numerical integration of the radial scattering equation, as implemented in GrayHawk, and one based on the WKB approximation. We consider four black hole geometries and scalar, fermionic, and vector perturbations, comparing the corresponding gray-body factors, primary Hawking emission spectra, and Page evaporation factors. As expected for high-frequency and large $l$ regimes, the two methods provide excellent agreement. At low frequencies, however, systematic discrepancies emerge and become particularly pronounced for nearly-extremal black holes. Systematic discrepancies are also met when considering effective geometric potentials characterized by the presence of more than one peak. A significant difference between the two approaches is also found in their computational cost, highlighting their complementary nature.
\end{abstract}

\section{Introduction}

The study of quantum fields propagating in Black Hole (BH) spacetimes provides a valuable framework for investigating the interplay between quantum field theory and curved spacetime. In particular, Hawking radiation \cite{Hawking:1975vcx} predicts that BHs emit particles with a thermal spectrum characterized by the Hawking temperature \cite{Gibbons:1977mu}. However, the spectrum observed by a distant observer is not purely thermal \cite{Kirchhoff01071860}. Particles emitted near the event horizon must propagate through the effective curvature potential generated by the BH geometry, which partially reflects the outgoing radiation. This effect is encoded in the Gray-Body Factors (GBFs), which determine the transmission probability of each field mode through the effective potential barrier \cite{Andersson:2000tf,Sakalli:2022xrb}. Consequently, an accurate computation of the GBFs is essential for obtaining reliable Hawking emission spectra and for determining quantities such as the energy and particle emission rates and the corresponding BH evaporation timescales.\\
It is important to remember that despite GBFs main application historically is in Hawking radiation and in the absorption cross section in scattering process \cite{Unruh:1976fm,Sanchez:1976fcl,Sanchez:1977si}, recently their connection with gravitational waves ringdown amplitudes \cite{Rosato:2024arw,Rosato:2026apq,Oshita:2023cjz,Okabayashi:2024qbz} and Quasi Normal Modes (QNMs) \cite{Konoplya:2024lir,Konoplya:2024vuj,Pedrotti:2025idg} was shown.\\
The computation of GBFs generally requires solving a scattering problem for the radial perturbation equation associated with the field under consideration \cite{Regge:1957td,Zerilli:1970se,Teukolsky:1973ha,Teukolsky:1972my,Press:1973zz,Teukolsky:1974yv,Page:1976df,Page:1976ki,Page:1977um}. For generic static, spherically symmetric BH geometries, the radial equation can be written in a Schr\"odinger-like form involving an effective potential whose structure depends on the background geometry, the spin of the field, and its angular momentum mode $l$ \cite{Chandrasekhar:1985kt,Kodama:2003kk,Kodama:2003jz}. The GBF is then identified with the transmission probability through this potential. A direct numerical integration of the radial equation provides a reliable approach that does not require additional approximations, but can become computationally demanding when a large number of frequencies, multipolar modes, or different BH configurations must be explored.\\
Semi-analytical approximations provide an alternative approach. Among these, the WKB approximation \cite{Schutz:1985km} is particularly suited to scattering problems characterized by localized potential barriers. The WKB method determines the transmission coefficient from the properties of the effective potential in the vicinity of its maximum and can be systematically improved by including higher-order corrections \cite{Matyjasek:2026yiu}. This makes the WKB approach an attractive option when fast computations are required. However, the accuracy of the WKB approximation is controlled by the semiclassical character of the scattering problem. In the regime of large angular momentum, the potential is dominated by the centrifugal contribution proportional to $l(l+1)$, and the corresponding WKB expansion is therefore naturally organized in powers of $1/l$ \cite{Bender:1999box}, and becomes increasingly accurate in the eikonal limit $l\gg1$.
Moreover, at high frequencies, a different mechanism contributes to the validity of the approximation. Since the local wavelength scales as $\lambda\sim 1/\omega$, sufficiently large frequencies correspond to wavelengths short compared with the characteristic length scale over which the effective potential varies. The potential therefore varies only weakly over a wavelength, which improves the validity of the WKB approximation. Conversely, at low frequencies the wavelength becomes comparable to, or larger than, the characteristic width of the potential barrier, and the local WKB description becomes less reliable. Furthermore, the applicability of the method ceases when the effective potential develops a structure with more complex shapes, such as the presence of a secondary peak, that cannot be adequately captured by a local expansion around its maximum \cite{Konoplya:2011qq,Konoplya:2025afm,Bonanno:2025ksz,Konoplya:2019hml,Konoplya:2017wot}.\\
The relevance of these limitations becomes particularly important for regular BH solutions and for configurations approaching extremality. Modifications of the BH geometry can significantly alter the effective potential governing field propagation and consequently affect both the GBFs and the resulting Hawking emission spectra. In nearly extremal configurations, the geometrical potential can become increasingly complex, potentially challenging the assumptions underlying the WKB approximation. \\
A systematic comparison with a fully numerical computation is therefore necessary to assess the range of validity of the approximation and to determine how its limitations propagate to physical observables derived from the GBFs.\\
In this work, we perform such a comparison taking into account two independent approaches for computing GBFs. The first is GrayHawk \cite{Calza:2025whq,Calza:2026wuf}, a fully numerical code that obtains the transmission probability by directly integrating the radial perturbation equation. This code have been largely used since its publication \cite{Han:2026txd,Fan:2026mdu,Han:2026fpn,Han:2025cal,Muniz:2025iqu,Pedrotti:2025idg,Yuan:2025eyi,Calza:2025mwn,Calza:2025yfm,Calza:2026csk}. The second is a code based on a high-order WKB approximation \cite{Matyjasek:2026yiu} and it also draw grate attention in the community \cite{Alonso-Bardaji:2026upq,Lutfuoglu:2026pgn,Konoplya:2019xmn,Toshmatov:2015wga,Skvortsova:2024msa,Abdullaev:2026cmt,Lutfuoglu:2026zxj,Bolokhov:2026uol,Ovgun:2026gac,Arbelaez:2026cnd} .\\
We consider different BH geometries as test cases, which can be divided into two metric macro-groups according to the different analyses we implemented:
\begin{itemize}
    \item Schwarzschild BH, Hayward BH \cite{Hayward:2005gi} and a solution inspired by Asymptotic Safety arguments\cite{Bonanno:2025dry}.
    \item A phenomenologically motivated modification of the Schwarzschild BH that yields a photonic effective potential with a multi-peak structure \cite{Konoplya:2024wds,Konoplya:2018yrp,Oshita:2024fzf}.
\end{itemize} 
Concerning the first metric macro-group, our analysis focuses on massless scalar, fermionic, and vector perturbations and compares the first three GBFs obtained with the two methods.
In the case of regular BH geometries, we investigate configurations with different values of the regularizing parameter.\\
Beyond the direct comparison of the GBFs, we investigate how the differences between the two computational approaches affect physical quantities derived from them. In particular, we compare the primary Hawking emission spectra and the Page evaporation factors, which encode the mass-loss rate associated with the emission of a field of a given spin \cite{Page:1976df,Page:1976ki}. This allows us to distinguish discrepancies that may be visually small at the level of individual GBFs but can become significant when integrated over the full frequency range \cite{Arbey:2025dnc}.\\
The analysis reveals two qualitatively distinct regimes. For non-near-extremal BH configurations, the WKB approximation reproduces the fully numerical results with high accuracy in the high-frequency and high-$l$ regimes, while deviations become increasingly relevant at low frequencies. For nearly extremal configurations, the discrepancies are substantially enhanced, particularly for higher-spin fields, and can significantly affect both the low-energy emission spectrum and the Page evaporation factors. The comparison therefore provides not only an assessment of the accuracy of the WKB approximation, but also a characterization of the regimes in which a fully numerical treatment is necessary.\\
Additionally, we compare the computational cost of the two approaches. The WKB method is substantially faster than the direct numerical computation, providing individual GBFs in a fraction of a second, whereas the GrayHawk calculation can require from tens of seconds to several minutes depending on the complexity of the metric. This difference makes the WKB approach particularly advantageous for rapid calculations in regimes where its accuracy is well established, especially at high energies. Conversely, the fully numerical GrayHawk approach, although computationally more demanding, retains its accuracy in the low-energy regime and is therefore better suited for calculations in which the low-frequency part of the spectrum or low multipole configurations play a significant role.\\
Finally, we consider the second macro-group. In such a case, a multiple-peak structure in the effective potential changes the dynamics of perturbations, forcing the wave to undergo successive scattering processes. The additional peak is a barrier off which the wave has to scatter, being transmitted or reflected. Therefore, between the two peaks, a partially confined cavity forms, and the wave undergoes repeated reflections before escaping or being absorbed,  generating delayed contributions to the waveform and commonly producing a modulation of the late-time signal referred to as echoes. The echoes' temporal separation depends on the distance between the peaks, and their amplitude depends on the transmission and reflection coefficients of the barriers.\\
This resonant structure has consequences for the perturbation dynamics, altering the QNM spectrum and the GBFs \cite{Berti:2025hly,Berti:2022xfj,Cardoso:2016rao,Cardoso:2017cqb,Rosato:2024arw}. Thus, a multi-peaked effective potential can lead to partial-wave trapping, repeated backscattering, and a characteristic echo structure in the evolution of perturbations. 
Studying the second metric macro-group we focused solely on the $s=1$ field and compared the GBFs obtained with the two methods, highlighting how the WKB method, pivoting around an approximated expansion of the geometric potential on the primary peak, lacks to capture the multiple reflections and transmissions that the wave undergoes between the two peaks, therefore lacking in describing the so-called echoes. \\
The remainder of this paper is organized as follows. In Sec. \ref{TheoreticalFormalism}, we introduce the theoretical framework for the propagation of massless fields in static, spherically symmetric BH spacetimes and define the relevant emission observables. In Sec. \ref{NumericalMethods}, we describe the two computational approaches employed in the analysis. The BH geometries considered in this work are reviewed in Sec. \ref{Metrics}. In Sec. \ref{Results}, we present and discuss the comparison between the two methods, focusing on the GBFs, primary emission spectra, and Page evaporation factors. Finally, in Sec. \ref{Conclusion}, we summarize our main findings and discuss the complementary roles of the two computational approaches.

\section{Theoretical formalism}
\label{TheoreticalFormalism}
In this section, we briefly review the theoretical framework employed throughout this work to compute GBFs and Hawking emission spectra for massless test fields propagating on generic static and spherically symmetric BH backgrounds.

\noindent We consider generic static,  spherically time-radial symmetric BH geometries described by the line element

\begin{equation}
ds^2=-f(r)dt^2+\frac{dr^2}{f(r)}+r^2\left(d\theta^2+\sin^2\theta\,d\phi^2\right),
\label{metric}
\end{equation}
where $f(r)$ is a function of the radial variable only and possesses an outermost root, such that $f(r_h)=0$ and $f(r)>0$ for all $r>r_h$, where the event horizon is located.
This class of metrics is such that the product of the coefficients of the $dt^2$ and $dr^2$ components is equal to $-1$, and the function multiplying the angular part is the areal radius when written in four-dimensional Boyer–Lindquist coordinates. 

\noindent It is possible to reduce the propagation of linear test perturbations on such a background into a single separable master equation \cite{Chandrasekhar:1985kt,Kodama:2003jz,Regge:1957td} whose radial part can be written as a one-dimensional Schr\"odinger-like wave equation. Introducing the tortoise coordinate

\begin{equation}
\frac{dr_*}{dr}=\frac{1}{f(r)},
\label{tortoise}
\end{equation}
the radial equation takes the form

\begin{equation}
\partial_{r_*}^2 \Psi_{s,l}
+\left(\omega^2-V_{s,l}(r)\right)\Psi_{s,l}=0,
\label{master}
\end{equation}
where $\omega$ denotes the wave frequency and $V_{s}(r)$ is the so-called effective potential, whose explicit form depends on the spin $s$, the field multipolar mode number $l$, and on the background geometry. We consider massless spin-0 and spin-1 bosons, and massless Dirac fields of spin $s=1/2$. For the bosonic cases, the potentials read \cite{Konoplya:2011qq,Berti:2009kk}:

\begin{equation}
V_{s,l}(r)=
f(r)
\left(
\frac{l(l+1)}{r^2}
+
(1-s^2)\frac{f'(r)}{r}
\right),
\label{bosonicpotential}
\end{equation}
The fermionic case can be expressed in terms of supersymmetric partner
potentials and reads: 
\begin{equation}
V_{1/2,l}(r)=
W^2(r)
\pm
f(r)\partial_r W(r).
\label{diracpotential}
\end{equation}
where
\begin{equation}
W(r)=
\frac{\lambda\sqrt{f(r)}}{r},
\end{equation}
and $\lambda=l+\frac{1}{2}$.
We notice that the $\pm$ choice leads to the same transmission probability and therefore to identical GBFs.

\noindent In the asymptotic regions, far away ($r_* \rightarrow \infty$) and at the event horizon ($r_* \rightarrow -\infty$), the potentials vanish, and therefore the solutions reduce to plane waves. Imposing purely ingoing boundary conditions \cite{Harmark:2007jy,Konoplya:2026qql}, the scattering problem can be set as follows
\begin{align}
    &\Psi_{s,l}(r_*)\sim
T_{s,l}\,e^{-i\omega r_*}\qquad,
\qquad\;\;\;\;\;\;\;\;\;\;\;\;
r_*\rightarrow-\infty,
\label{horBC}\\
&\Psi_{s,l}(r_*)\sim
e^{-i\omega r_*}
+
R_{s,l}\,
e^{i\omega r_*}\qquad,
\qquad
r_*\rightarrow+\infty,
\label{infBC}
\end{align}
where $R_{s,l}$ and $T_{s,l}$ denote the reflection and transmission amplitudes, respectively, depends on the field spin $s$, mode $l$, frequency $\omega$, and satisfy
\begin{equation}
|R_{s,l}|^2+|T_{s,l}|^2=1.
\label{unitarity}
\end{equation}

\noindent The transmission probability $|T_{l s}|^2$ is often dubbed GBF and denoted with $\Gamma_{l s}(\omega)$.
It describes the probability that a particle emitted near the event horizon successfully propagates through the curvature potential, reaching an observer asymptotically distant. Deviations of $\Gamma_{l s}(\omega)$ from unity quantify deviations from an otherwise BH blackbody emission spectrum.

In our analysis, we assumed BHs in thermal equilibrium with their surroundings. Therefore, the spectral energy emission rate for a massless field of spin $s$ is given by \cite{Hawking:1974rv,Hawking:1975vcx,Page:1976df,Page:1976ki}

\begin{equation}
\frac{d^2E}{dtd\omega}
=
\frac{1}{2\pi}
\sum_{l}
g_{l} n_i
\frac{\omega\,\Gamma_{s,l}(\omega)}
{e^{\omega/T_H }\pm 1},
\label{EnergySpectrum}
\end{equation}
where $n_i$ denotes the number of internal degrees of freedom of the field. The plus (minus) sign corresponds to fermions (bosons), $g_{l}=2l+1$ denotes the degeneracy factor of each angular mode, and $T_H$ is the Hawking temperature given by \cite{Hawking:1975vcx,Gibbons:1977mu}:
\begin{equation}
T_H = \frac{f'(r)}{4\pi} \Big|_{r = r_H}.
\end{equation}  Likewise, the particle emission spectrum reads

\begin{equation}
\frac{d^2N}{dtd\omega}
=
\frac{1}{2\pi}
\sum_{l}
g_{l} n_i
\frac{\Gamma_{s,l}(\omega)}
{e^{\omega/T_H} \pm 1}.
\label{ParticleSpectrum}
\end{equation}

\noindent Enforcing the conservation of energy it is possible to conclude that the particle emission must come at the expense of the BH mass.
We follow the procedure described in \cite{Page:1976df,Page:1976ki,Page:1977um,Chambers:1997ai,Chambers:1997ax,Taylor:1998dk} to compute the energy extraction of each field $l$-mode of spin $s$, and summing over all modes we obtain the rates of mass loss due to the emission of a single specific field of spin-$s$, $f_s$:
\begin{equation}\label{f_s}
f_s
=\sum_{l}\frac{g_l}{2\pi} \int_0^{\infty}dx \frac{^i\Gamma_{s,l}}{e^{\omega/T_H} \pm 1}x~,
\end{equation}
where $x=\omega M$ and we removed the dependence on the BH mass.
One may notice that since we are considering massless fields, the integral in \eqref{f_s} spans all energies going from 0 to infinity.

\section{Numerical methods}
\label{NumericalMethods}

In this section, we describe the basic concepts and approximations characterizing two numerical methods taken into account and the two publicly available codes we employed for calculating the GBFs.

\subsection{GrayHawk}

\texttt{GrayHawk} \cite{Calza:2025whq} (and its subsequent implementation \texttt{GrayHawk} \texttt{v2} \cite{Calza:2026wuf}) is a Mathematica-based numerical code designed to solve the scattering problem of massless fields propagating on general static, spherically symmetric, and asymptotically flat BH spacetimes. It is applicable to fields of spin $s=0,1/2,1,2$ and does not rely on a WKB or other semi-analytical approximation. Instead, the radial wave equation is solved numerically, with the GBFs obtained directly from the transmission coefficient of the corresponding scattering problem. The code is formulated for metrics of the form
\begin{equation}
ds^2=-G(r)dt^2+\frac{dr^2}{F(r)}
+H(r)d\Omega^2,
\end{equation}
where asymptotic flatness requires
\begin{equation}
F(r)\rightarrow1,\qquad
G(r)\rightarrow1,\qquad
H(r)\sim r^2
\qquad (r\rightarrow\infty).
\end{equation}
The code therefore naturally embeds metrics of the form \eqref{metric}.

\noindent The numerical procedure implemented in \texttt{GrayHawk} is:

\begin{itemize}
    \item A spin $s$ for the field is seleted as well as a value for the angular momentum number $l$ and the radial perturbation equation is considered in its Schrödinger-like form \eqref{master}.
    \item It is selected a set of frequencies $\omega_i$ for which the computation will performed.  
    \item It is chosen an adequately close to the event horizon value to be aproximable as plane waves, where, and purely ingoing boundary conditions, and normalization of the wavefunction are imposed.
    \item For each $\omega_i$ a numerical integration outwards towards using \eqref{master} is performed up to an adequately large distance where the effective potential becomes negligible, and the solution is fitted with \eqref{infBC}.
    \item The value of the GBF at the specific $\omega_i$ is obtained given the ratio of the asymptotic amplitudes.
    \item The procedure is repeated for each value of $\omega_i$ required in the calculation, yielding the corresponding GBF.
\end{itemize}

\subsection{WKB code}

One of the most widely used approximate techniques when dealing with potential barriers having a constant value in the asymptotic regions is certainly the WKB approximation \cite{Schutz:1985km}.

\noindent The WKB formula is based on a matching procedure between the asymptotic solutions, expressed as a superposition of plane waves, and the solution in the region close to the maximum of the potential barrier. In particular, assuming that the effective potential has a single maximum at $r_m$, it can be expanded in a Taylor series around its maximum. Defining the two turning points by
\begin{equation}
V_{s,l}(r_*)=\omega^2,
\end{equation}
and matching the solutions across these turning points provides the results through the WKB approximation, allowing one to relate the amplitudes of the ingoing and outgoing waves and, consequently, to determine the reflection and transmission coefficients \cite{Konoplya:2011qq}.

\noindent Throughout the years, higher and higher orders in the expansion have been considered (up to the third order in \cite{Iyer:1986np}, up to the sixth order in \cite{Konoplya:2003ii}, and through thirteen in \cite{Matyjasek:2017psv}). The general expression for the frequencies is formulated as an expansion around the eikonal regime with this form \cite{Konoplya:2019hlu}:
\begin{align} \label{WKBord}
    &
    \omega^2 = \mathcal{V}_0 + A_2(K^2) + A_4(K^2) + A_6(K^2) + \ldots \nonumber 
    \\
&- iK\sqrt{-2V_2} \left[1 + A_3(K^2) + A_5(K^2) + A_7(K^2) + \ldots \right],
\end{align}
In this expression, $ \mathcal{V}_0 $ denotes the value of the effective potential peak of \eqref{master}, located at $r_m$, while
\begin{equation}
\mathcal{V}_n=\left.\partial^n_{r^*} V_{s,l}\right|_{r_*=r_m}
\end{equation}
denotes the $n$-th derivative of the effective potential with respect to the tortoise coordinate, evaluated at the maximum. In particular, $\mathcal{V}_2$ is the second derivative at the maximum. The coefficients $A_n$ ($n=2,3,4,\ldots$) are higher-order correction terms, which are polynomials of the derivatives $\mathcal{V}_2,\mathcal{V}_3,\ldots,\mathcal{V}_{2n}$ and of $K^2$\cite{Konoplya:2019hlu}.The quantity $K$ entering Eq.~\eqref{WKBord} can be defined as \cite{Konoplya:2003ii}

\begin{equation}
K = \frac{i\left(\omega^2 - V_{s,l}(r_m)\right)}{\sqrt{-2 \frac{d^2}{d r_*^2} V_{s,l}(r_m)}} + \sum_{p=2}^6 \Lambda_p(\omega,\mathcal{V}_n,K).
\label{Kwkb}
\end{equation}
Here, $\Lambda_p(\omega,\mathcal{V}_n,K)$ are higher-order WKB correction terms which depend on the frequency $ \omega$, on the potential and its derivatives evaluated at the peak $r_m$, and on $K$. Therefore, the equation must be solved numerically with respect to $K$. Specifically, to compute the GBFs throughout this paper, we used the sixth-order WKB expansion, including the correction terms $\Lambda_p$ up to $p=6$ whose explicit expressions can be found in \cite{Konoplya:2003ii}. At leading order, Eq.~\eqref{WKBord} provides the unique solution
\begin{equation}
K_{\rm eik}=
\frac{i\left(\omega^2-\mathcal{V}_0\right)}
{\sqrt{-2\mathcal{V}_2}}.
\end{equation}
When higher-order corrections are included, the resulting equation \eqref{Kwkb} has multiple solutions for $K$ at a fixed frequency $\omega$. The solution corresponding to the perturbative correction of the eikonal result is selected as the root closest to $K_{\rm eik}$ \cite{Konoplya:2019hlu}.

\noindent In the case of a real effective potential, $K$ is a purely imaginary quantity \cite{Iyer:1986np} and is related to the reflection and transmission coefficients as
\begin{equation}
|R|^2 = \frac{1}{1 + e^{-2 i \pi K}}, \quad |T|^2 = \Gamma_{l s}= \frac{1}{1 + e^{2 i \pi K}},
\end{equation}
where $K$ is given by \eqref{Kwkb}. Therefore, the GBF is obtained as the transmission probability through the effective potential barrier.

\noindent The implementation follows the automatic WKB procedure described in \cite{Konoplya:2019hlu}, which allows the GBFs to be obtained directly from the effective potential and its derivatives at the maximum. The corresponding Mathematica code is publicly available in the repository \cite{link} associated with \cite{Matyjasek:2026yiu}. By providing the metric function and the corresponding effective potential, the GBFs can be calculated automatically for the different fields and multipolar modes considered in this work.

\section{Metrics}
\label{Metrics}
In this section, we propose a brief review of the metrics we employed as test geometries in the analysis we will outline in the next section.

\subsection{Schwarzschild}
The Schwarzschild spacetime is the simplest exact vacuum solution of Einstein's equation. It was discovered by Karl Schwarzschild in 1916 \cite{Schwarzschild:1916uq} and represents a static, spherically symmetric BH with no charge nor rotation, and is given by a line element of the form of \eqref{metric} with metric function
\begin{equation}
    f(r) = 1 - \frac{2M}{r}
\end{equation}
where \(M\) is the mass of the BH.  
The Schwarzschild BH has a single event horizon located at \(r_H = 2M\). This is the radius at which the escape velocity equals the speed of light, effectively trapping all matter and radiation. 
This spacetime solution contains an irremovable singularity in \(r = 0\), where the spacetime curvature becomes infinite.

\subsection{Hayward metric}

Originally proposed as a simple model of a singularity-free spacetime, the Hayward metric describes a static, spherically symmetric, and asymptotically flat spacetime, where the central curvature singularity is replaced by a regular de Sitter core \cite{Hayward:2005gi}.

\noindent Its metric function is given by

\begin{equation}
f(r)=
1-\frac{2Mr^2}{r^3+2M \ell^2},
\label{HaywardMetric}
\end{equation}
where $M$ denotes the BH mass and $\ell$ is the so-called regularizing parameter.

\noindent For $\ell\rightarrow0$, Eq.~(\ref{HaywardMetric}) reduces smoothly to the Schwarzschild case, and for non-vanishing values of $\ell$ near the origin the metric function behaves as

\begin{equation}
f(r)\simeq
1-\frac{r^2}{\ell^2},
\qquad
r\rightarrow0,
\end{equation}
corresponding to a de Sitter core and ensuring that all curvature invariants
remain finite.

\noindent Depending on the value of the dimensionless parameter $\ell/M$, the spacetime may
possess two horizons (of which the innermost is a Cauchy horizon subjected to mass inflation instabilities), a degenerate extremal horizon, or no horizons if, respectively, $\ell/M\lesseqgtr\sqrt{16/27}$. In the present work we restrict our analysis to the BH regime, considering only values of $\ell$ for which an event horizon exists.

\subsection{Asymptotic safety metric}

Within the Asymptotic Safety framework \cite{Weinberg:1979,Reuter:1996cp,Niedermaier:2006wt} were developed several static solutions \cite{Spina:2025wxb,Bonanno:2000ep,Platania:2019kyx,Held:2019xde} and, through the proper-time flow approach \cite{deAlwis:2017ysy,Bonanno:2020bil,Bonanno:2025tfj}, it is possible to draw a dynamical gravitational collapse scenario where quantum corrections modify the gravitational interaction during the collapse \cite{Bonanno:2025dry,Bonanno:2023rzk,Harada:2025cwd}.

\noindent Matching the collapsing matter \cite{Markov1985DeSI} with an exterior standard static and spherically symmetric geometry \cite{Israel:1966rt,Malafarina:2022wmx}, it is possible to obtain a BH solution having the following metric function  \cite{Bonanno:2025dry} \footnote{We note that in the previous works \cite{Bonanno:2025dry} the metric function contains a typo. Namely, the mass $M$ in the first square root is not reported squared. We also notice that such a typo does not influence the computations or conclusions therein drawn since the mass parameter is normalized to 1.}

\begin{equation} \label{metricAS}
    f(r)=\frac{3M^2+qr^4-M\sqrt{9M^2+q^2 r^6}}{qr^4}+\frac{2}{3}qr^2\text{Arctanh}\left(\frac{\left(q-\sqrt{q^2+\frac{9M^2}{r^6}}\right)r^3}{3M}\right)
\end{equation} 
with $M$ the asymptotic BH mass and $q$ encoding the quantum gravity corrections originating from the proper-time renormalization group flow.

\noindent In the classical limit, corresponding to $q\rightarrow\infty$, the metric
function reduces to the Schwarzschild solution, and the expansion for $r\gg q$ gives the Schwarzschild lapse function plus higher-order corrections controlled by inverse powers of $q$. 

\noindent The horizon structure depends on the value of the parameter
$q$. The solution admits inner and outer horizon, a degenerate horizon or no horizon when $q>q_c$, $q=q_c$ or $q<q_c$, respectively, where $q_{cr}$ is the critical value which is approximately $q_{\rm cr}\sim1.37/M^2$. 
When $q>q_c$ the the geometry at the origin is characterized by a regular core.

\subsection{Phenomenology-inspired geometry with a multi-peaked effective potential}
\label{subsec:multi_peaked}
To extend our analysis to configurations beyond the metrics characterized by a single peak in the geometric effective potential, we consider a phenomenological modification of the Schwarzschild metric. We realize it by modifying the electromagnetic ($s=1$) geometric potential of the Schwarzschild spacetime, introducing a Gaussian perturbation characterized by an height $\eta$, a width $\sigma$, and a central position $r_e$. Doing so, we introduce an additional peak in the effective potential to investigate how the fully numerical and WKB methods respond to a configuration where the wave propagation occurs in the presence of additional localized structures in the scattering potential. Those types of deviations from the Schwarzschild geometry can account for the presence of surrounding matter or environmental effects \cite{Konoplya:2025ixm,Cardoso:2021wlq,Macedo:2015ikq} and are sometimes dubbed dirty BHs.
When focusing on this specific metric, we restrict our analysis to the photon field and to construct its line element we start by considering the effective potential of the $s=l=1$ mode and add the perturbation 
\begin{equation}
V_{1,1}(r) = V_{1,1}^{\rm Sch}(r) + \delta V(r),
 \label{eq:multi_peak_potential}
\end{equation}
where from Eq. (\ref{bosonicpotential}) \begin{equation}
    V_{1,1}^{\rm Sch}(r)=\frac{2}{r^2} \left(1-\frac{2M}{r}\right)
\end{equation}
and $\delta V(r)$ parametrizes the additional structure,
\begin{equation}
    \delta V(r)
    = \eta \left(1-\frac{2M}{r}\right)
      \exp\left[-\frac{(r-r_e)^2}{2\sigma^2}\right],
    \label{eq:gaussian_perturbation}
\end{equation}
We then reconstruct the metric function associated with the modified potential \eqref{eq:multi_peak_potential} in the case $s=l=1$, using \eqref{bosonicpotential}, obtaining
\begin{equation}
    f(r) =
    \left(1-\frac{2M}{r}\right)
    \left[
        1+\frac{\eta r^2}{2}
        \exp\left(-\frac{(r-r_e)^2}{2\sigma^2}\right)
    \right].
    \label{eq:multi_peak_metric}
\end{equation}\\
Once the metric is fixed, we use the same geometry to compute the GBFs for electromagnetic perturbations with $l=1,2,3$.
In realistic astrophysical scenarios, this additional structure accounts for environmental matter surrounding the BH, for example, the presence of an accretion disk \cite{Page:1974he,Abramowicz:2011xu,Novikov:1973kta}, which is typically located at $r_e \sim (10-50)M$. \\
In the next section, we will consider a Gaussian term such that the resulting effective potential exhibits a second peak, smaller and distinct with respect to the main barrier.
This setup will allow us to compare numerical and sixth-order WKB calculations of the GBFs in a configuration where the effective potential has a more complex and physically motivated structure with respect to the standard barrier provided by vacuum solutions.

\section{Results and Comparison}
\label{Results}

\begin{figure}[H]
    \centering

    \begin{subfigure}[b]{0.46\textwidth}
        \centering
        \includegraphics[width=0.9\textwidth]{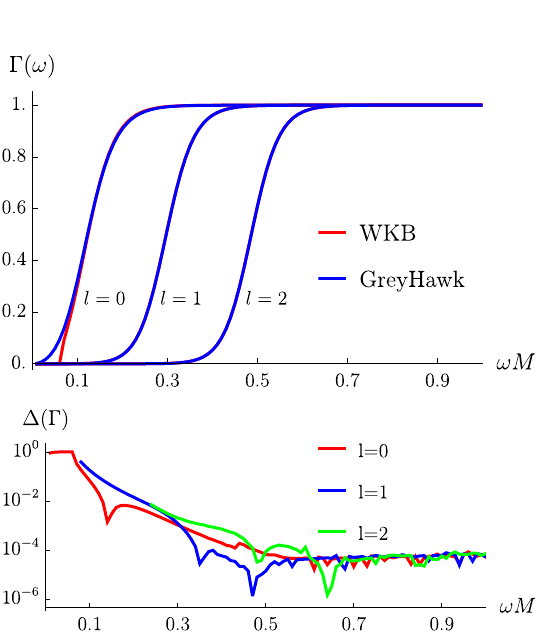}
        \caption{Scalar field.}
        \label{fig:scalar}
    \end{subfigure}
    \hfill
    \begin{subfigure}[b]{0.46\textwidth}
        \centering
        \includegraphics[width=0.9\textwidth]{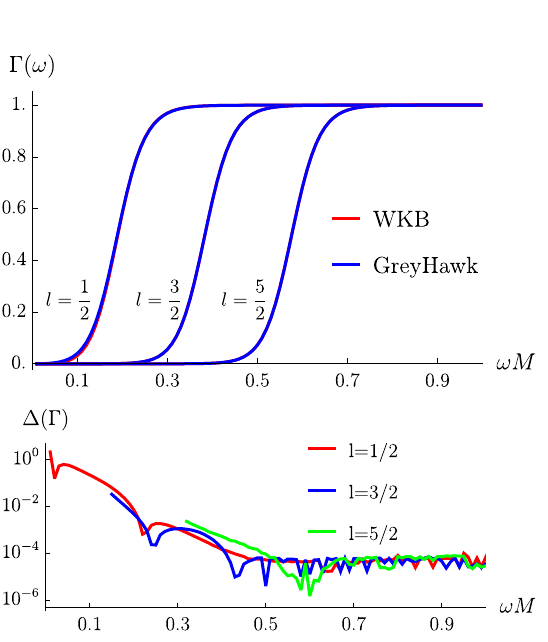}
        \caption{Fermions.}
        \label{fig:fermions}
    \end{subfigure}
    \hfill    \begin{subfigure}[b]{0.46\textwidth}
        \centering
        \includegraphics[width=0.9\textwidth]{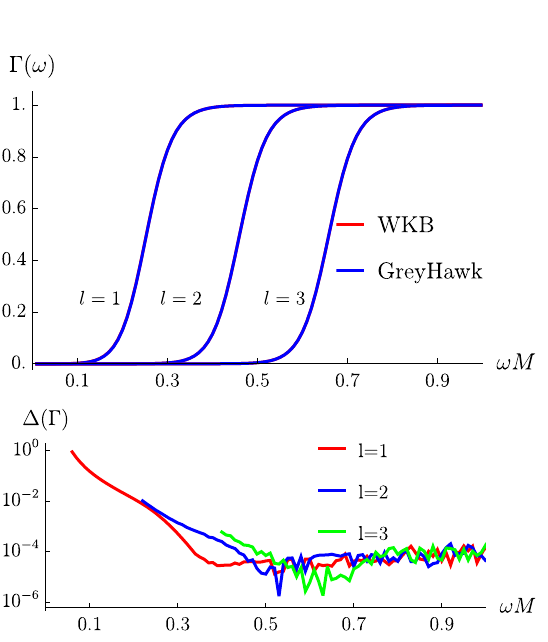}
        \caption{Photon.}
        \label{fig:photon}
    \end{subfigure}
    \begin{subfigure}[b]{0.46\textwidth}
        \centering
        \includegraphics[width=\textwidth]{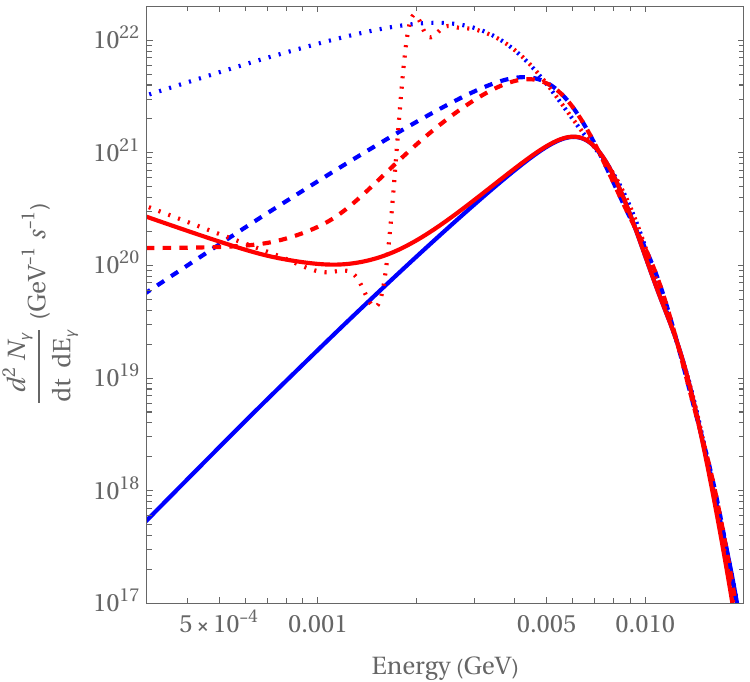}
        \caption{Spectrum}
        \label{fig:sch}
    \end{subfigure}

    \caption{This figure shows results for a Schwarzschild BH. Panels (a), (b), and (c) show the GBFs and corresponding $\Delta(\Gamma)$, as defined in (\ref{DeltaGamma}) for massless fields of spin $s=0,1/2,1$ respectively. Panel (d) shows the primary spectra emitted by a BH of mass $M=10^{13}$ kg for $s=0,1/2,1$ fields, denoted in dotted, dashed, and solid lines respectively. All plots, except the $\Delta(\Gamma)$ ones, share the same color code: in blue are reported the results computed using GrayHawk, while the 6th-order WKB method is reported in red. The plots showing the quantity $\Delta(\Gamma)$ report in red, blue, and green, the values of $\Delta(\Gamma)$ for the primary emitted mode of each field, namely the one with the lowest possible $l$, and the next two modes, respectively. Those last plots are cut when the values of the GrayHawk-computed GBF drop below $10^{-4}$.}
    \label{fig:complete}
\end{figure}

\begin{figure}[H]
    \centering

    \begin{subfigure}[b]{0.46\textwidth}
        \centering
        \includegraphics[width=0.9\textwidth]{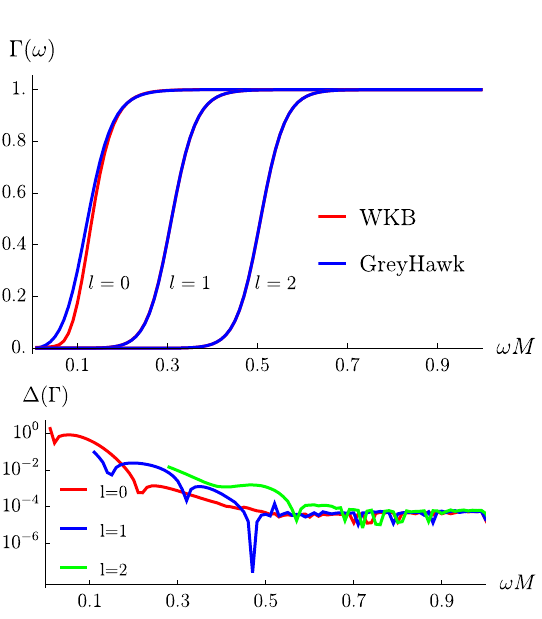}
        \caption{Scalar field.}
        \label{fig:scalarHay}
    \end{subfigure}
    \hfill
    \begin{subfigure}[b]{0.46\textwidth}
        \centering
        \includegraphics[width=0.9\textwidth]{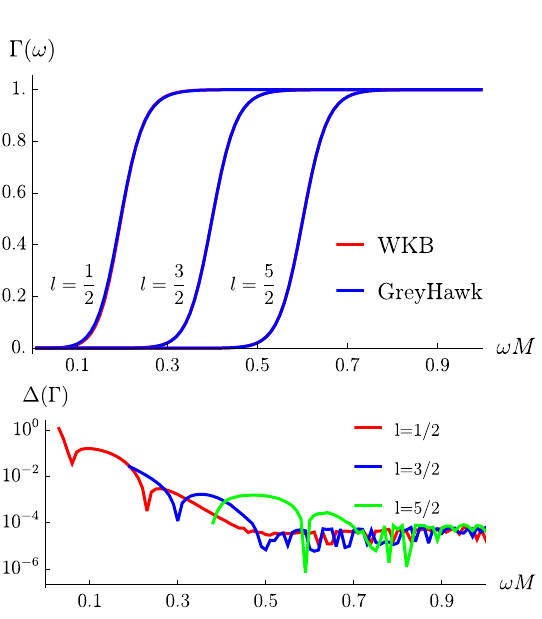}
        \caption{Fermions.}
        \label{fig:fermionsHay}
    \end{subfigure}
    \hfill

     \begin{subfigure}[b]{0.46\textwidth}
        \centering
        \includegraphics[width=0.9\textwidth]{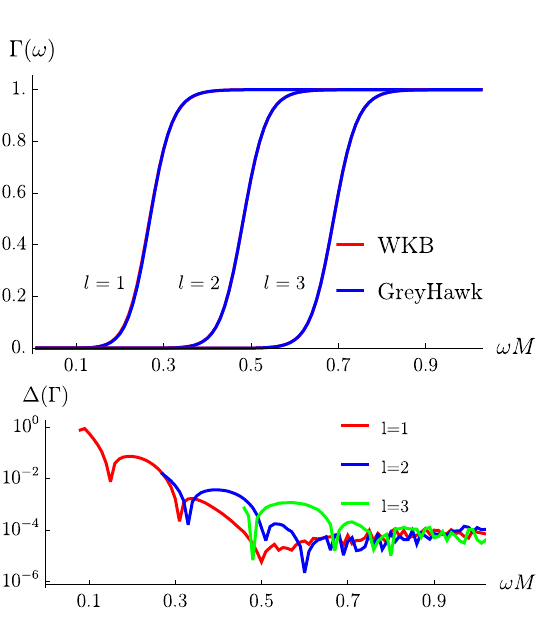}
        \caption{Photon.}
        \label{fig:photonHay}
    \end{subfigure}
    \begin{subfigure}[b]{0.48\textwidth}
        \centering
        \includegraphics[width=\textwidth]{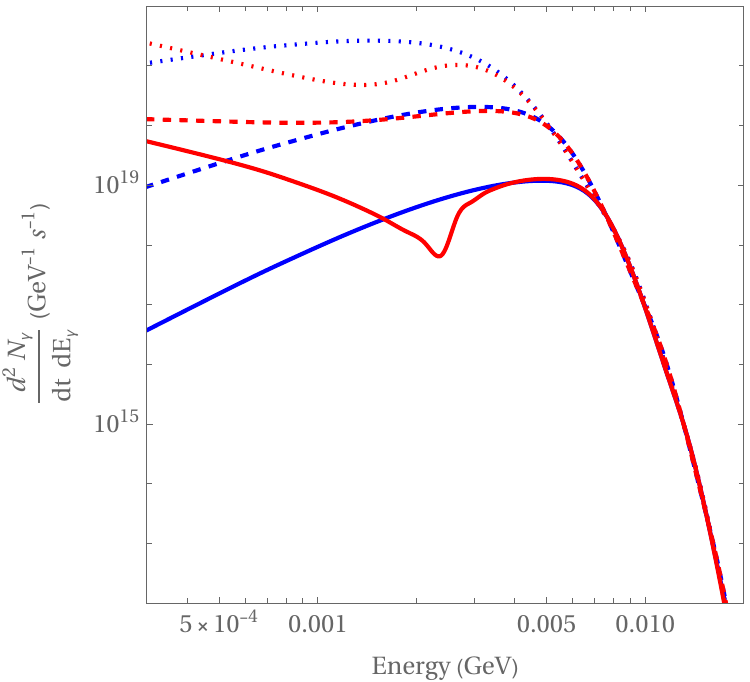}
        \caption{Spectrum}
        \label{fig:Hay}
    \end{subfigure}

    \caption{This figure shows results for an H09 BH. Panels (a), (b), and (c) show the GBFs and corresponding $\Delta(\Gamma)$, as defined in (\ref{DeltaGamma}) for massless fields of spin $s=0,1/2,1$ respectively. Panel (d) shows the primary spectra emitted by a BH of mass $M=10^{13}$ kg for $s=0,1/2,1$ fields, denoted in dotted, dashed, and solid lines respectively. All plots, except the $\Delta(\Gamma)$ ones, share the same color code: in blue are reported the results computed using GrayHawk, while the 6th-order WKB method is reported in red. The plots showing the quantity $\Delta(\Gamma)$ report in red, blue, and green, the values of $\Delta(\Gamma)$ for the primary emitted mode of each field, namely the one with the lowest possible $l$, and the next two modes, respectively. Those last plots are cut when the values of the GrayHawk-computed GBF drop below $10^{-4}$.}
    \label{fig:completeHay}
\end{figure}

\begin{figure}[H]
    \centering

    \begin{subfigure}[b]{0.46\textwidth}
        \centering
        \includegraphics[width=0.9\textwidth]{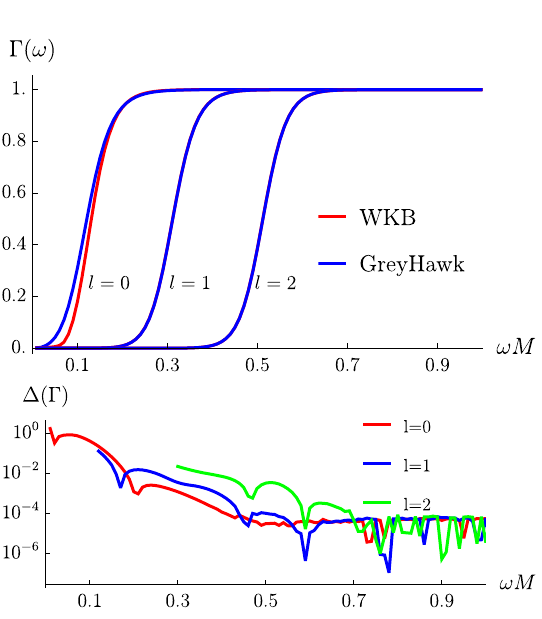}
        \caption{Scalar field.}
        \label{fig:scalarHay99}
    \end{subfigure}
    \hfill
    \begin{subfigure}[b]{0.46\textwidth}
        \centering
        \includegraphics[width=0.9\textwidth]{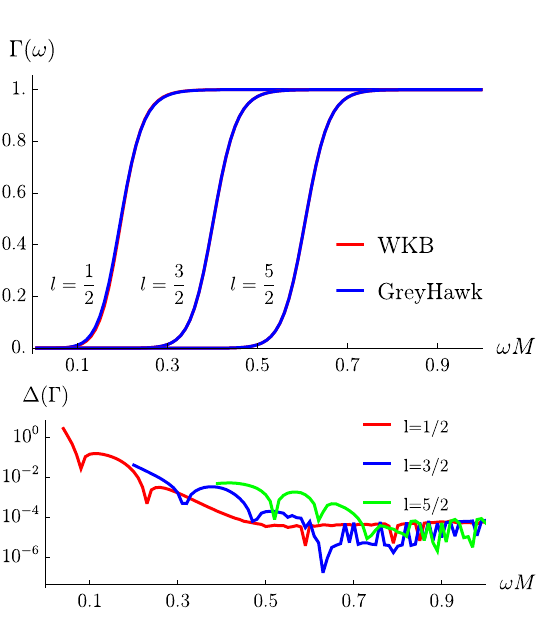}
        \caption{Fermions.}
        \label{fig:fermionsHay99}
    \end{subfigure}
    \hfill

      \begin{subfigure}[b]{0.46\textwidth}
        \centering
        \includegraphics[width=0.9\textwidth]{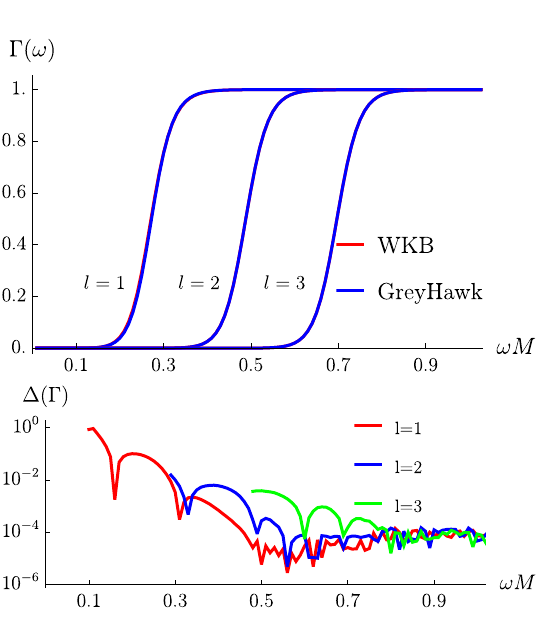}
        \caption{Photon.}
        \label{fig:photonHay99}
    \end{subfigure}
    \begin{subfigure}[b]{0.48\textwidth}
        \centering
        \includegraphics[width=\textwidth]{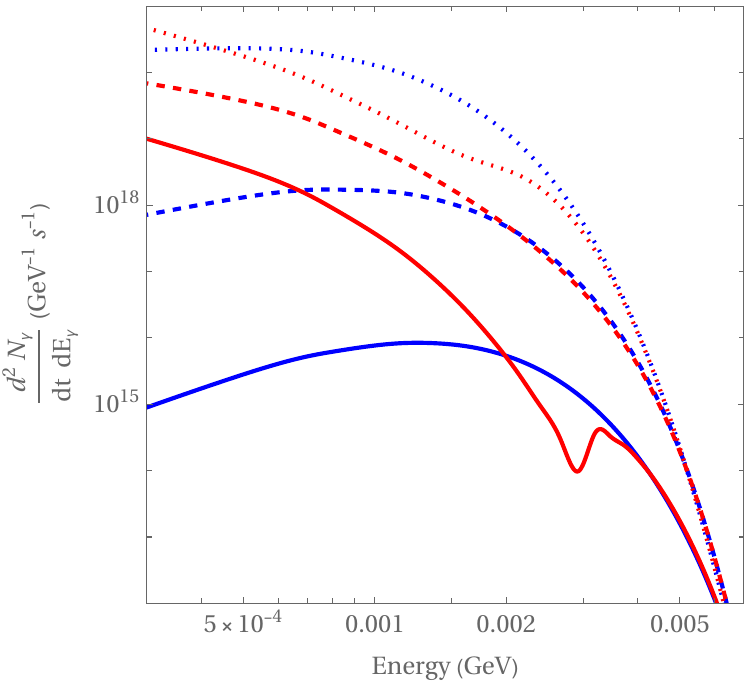}
        \caption{Spectrum}
        \label{fig:Hay99}
    \end{subfigure}

    \caption{This figure shows results for an H99 BH. Panels (a), (b), and (c) show the GBFs and corresponding $\Delta(\Gamma)$, as defined in (\ref{DeltaGamma}) for massless fields of spin $s=0,1/2,1$ respectively. Panel (d) shows the primary spectra emitted by a BH of mass $M=10^{13}$ kg for $s=0,1/2,1$ fields, denoted in dotted, dashed, and solid lines respectively. All plots, except the $\Delta(\Gamma)$ ones, share the same color code: in blue are reported the results computed using GrayHawk, while the 6th-order WKB method is reported in red. The plots showing the quantity $\Delta(\Gamma)$ report in red, blue, and green, the values of $\Delta(\Gamma)$ for the primary emitted mode of each field, namely the one with the lowest possible $l$, and the next two modes, respectively. Those last plots are cut when the values of the GrayHawk-computed GBF drop below $10^{-4}$.}
    \label{fig:completeHay99}
\end{figure}

\begin{figure}[H]
    \centering

    \begin{subfigure}[b]{0.46\textwidth}
        \centering
        \includegraphics[width=0.9\textwidth]{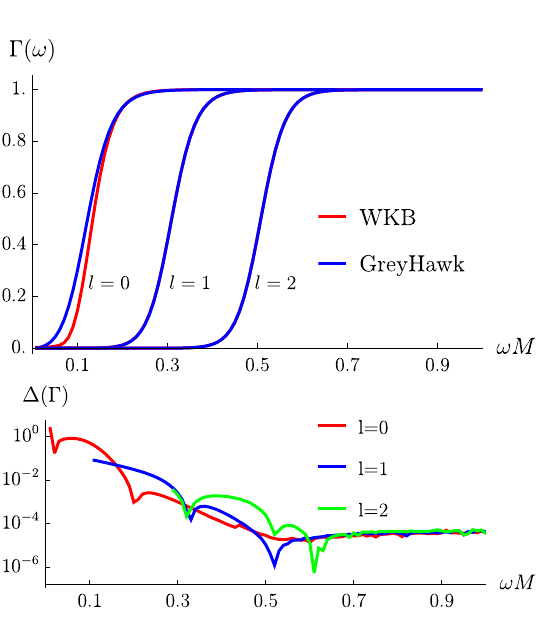}
        \caption{Scalar field.}
        \label{fig:scalarAS}
    \end{subfigure}
    \hfill
    \begin{subfigure}[b]{0.46\textwidth}
        \centering
        \includegraphics[width=0.9\textwidth]{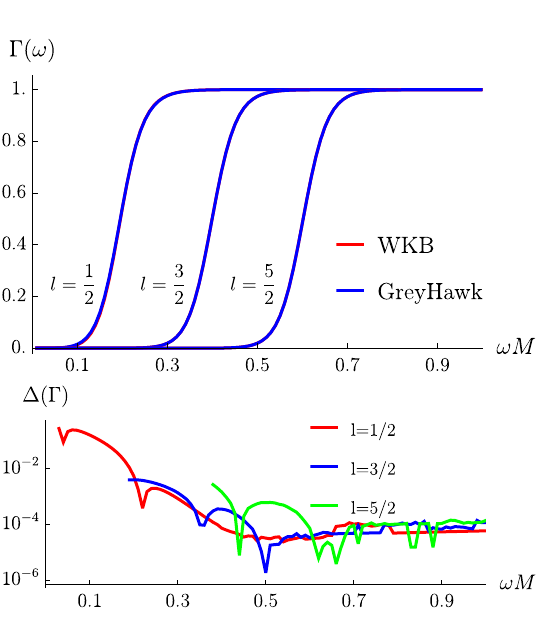}
        \caption{Fermions.}
        \label{fig:fermionsAS}
    \end{subfigure}
    \hfill

    \begin{subfigure}[b]{0.46\textwidth}
        \centering
        \includegraphics[width=0.9\textwidth]{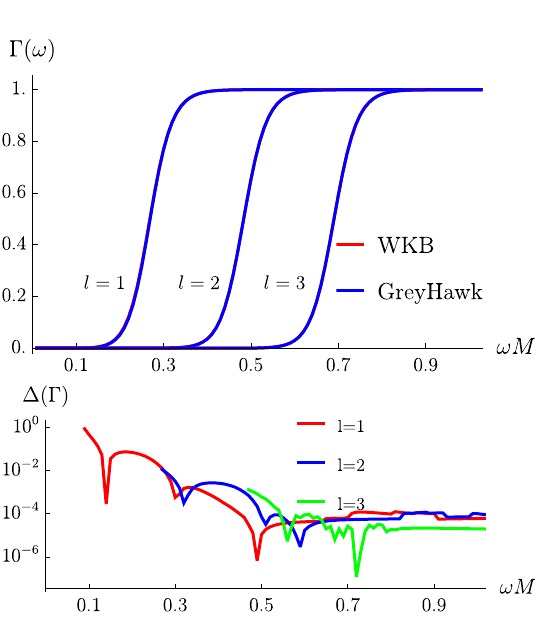}
        \caption{Photon.}
        \label{fig:photonAS}
    \end{subfigure}
    \begin{subfigure}[b]{0.48\textwidth}
        \centering
        \includegraphics[width=\textwidth]{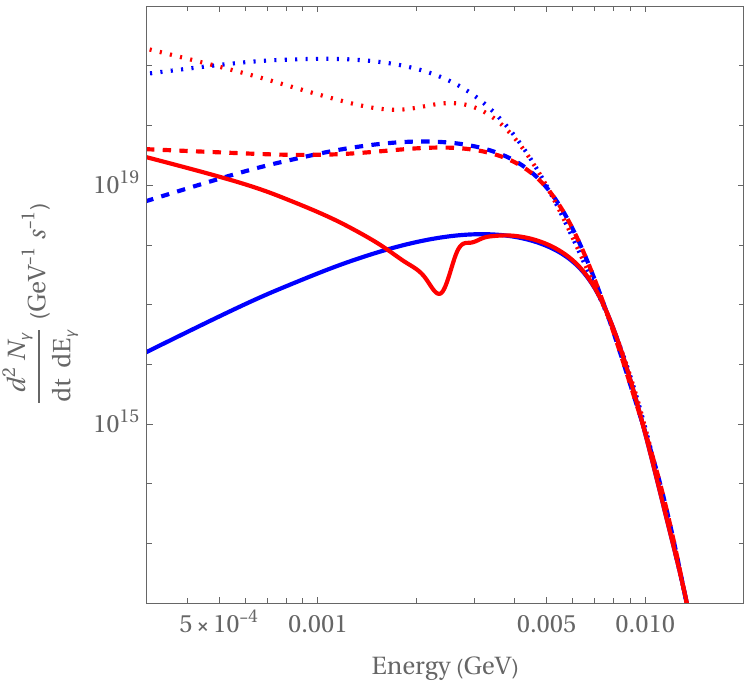}
        \caption{Spectrum}
        \label{fig:AS}
    \end{subfigure}

    \caption{This figure shows results for an AS09 BH. Panels (a), (b), and (c) show the GBFs and corresponding $\Delta(\Gamma)$, as defined in (\ref{DeltaGamma}) for massless fields of spin $s=0,1/2,1$ respectively. Panel (d) shows the primary spectra emitted by a BH of mass $M=10^{13}$ kg for $s=0,1/2,1$ fields, denoted in dotted, dashed, and solid lines respectively. All plots, except the $\Delta(\Gamma)$ ones, share the same color code: in blue are reported the results computed using GrayHawk, while the 6th-order WKB method is reported in red. The plots showing the quantity $\Delta(\Gamma)$ report in red, blue, and green, the values of $\Delta(\Gamma)$ for the primary emitted mode of each field, namely the one with the lowest possible $l$, and the next two modes, respectively. Those last plots are cut when the values of the GrayHawk-computed GBF drop below $10^{-4}$.}
    \label{fig:completeAS}
\end{figure}

\begin{figure}[H]
    \centering

    \begin{subfigure}[b]{0.46\textwidth}
        \centering
        \includegraphics[width=0.9\textwidth]{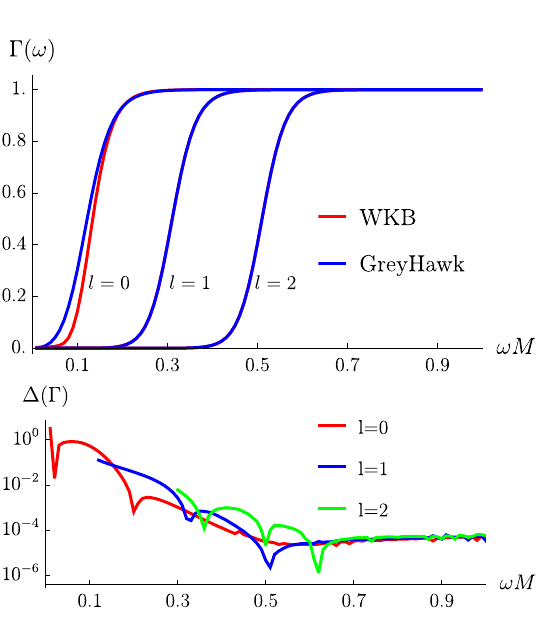}
        \caption{Scalar field.}
        \label{fig:scalarAS99}
    \end{subfigure}
    \hfill
    \begin{subfigure}[b]{0.46\textwidth}
        \centering
        \includegraphics[width=0.9\textwidth]{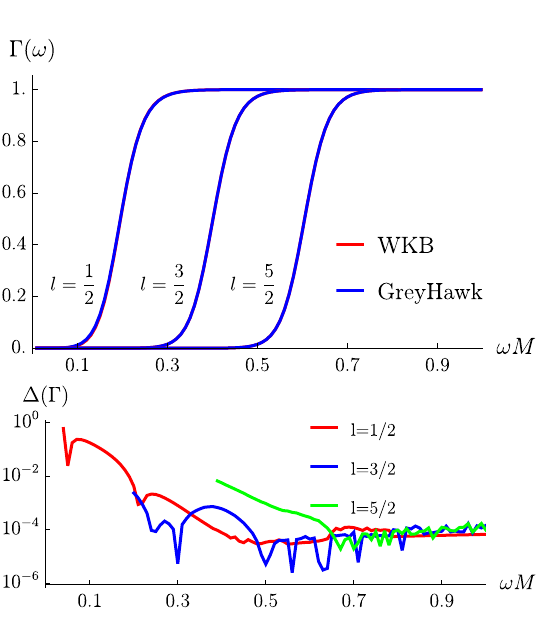}
        \caption{Fermions.}
        \label{fig:fermionsAS99}
    \end{subfigure}
    \hfill

    \begin{subfigure}[b]{0.46\textwidth}
        \centering
        \includegraphics[width=0.9\textwidth]{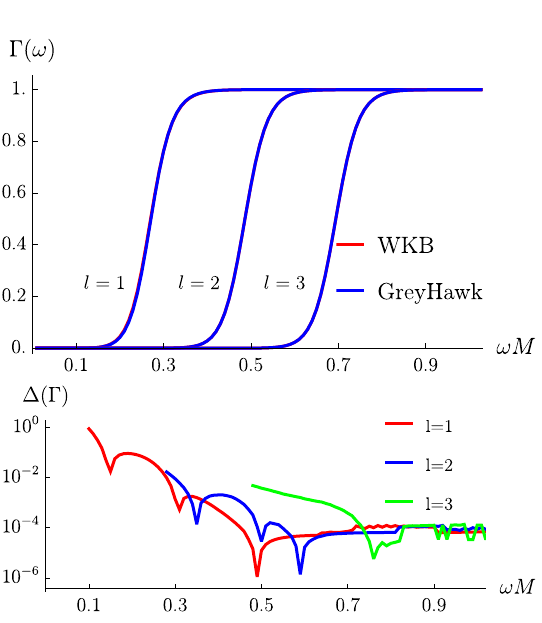}
        \caption{Photon.}
        \label{fig:photonAS99}
    \end{subfigure}
    \begin{subfigure}[b]{0.48\textwidth}
        \centering
        \includegraphics[width=\textwidth]{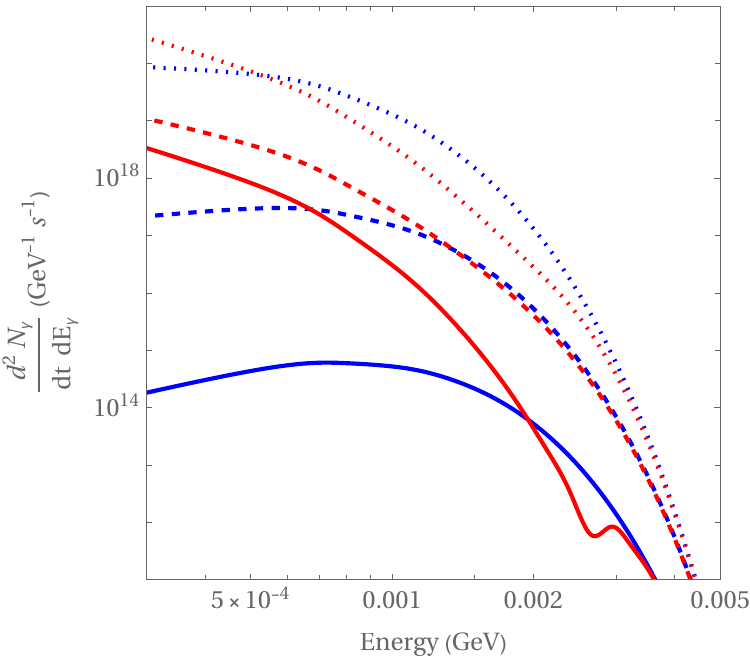}
        \caption{Spectrum}
        \label{fig:AS99}
    \end{subfigure}

    \caption{This figure shows results for an AS99 BH. Panels (a), (b), and (c) show the GBFs and corresponding $\Delta(\Gamma)$, as defined in (\ref{DeltaGamma}) for massless fields of spin $s=0,1/2,1$ respectively. Panel (d) shows the primary spectra emitted by a BH of mass $M=10^{13}$ kg for $s=0,1/2,1$ fields, denoted in dotted, dashed, and solid lines respectively. All plots, except the $\Delta(\Gamma)$ ones, share the same color code: in blue are reported the results computed using GrayHawk, while the 6th-order WKB method is reported in red. The plots showing the quantity $\Delta(\Gamma)$ report in red, blue, and green, the values of $\Delta(\Gamma)$ for the primary emitted mode of each field, namely the one with the lowest possible $l$, and the next two modes, respectively. Those last plots are cut when the values of the GrayHawk-computed GBF drop below $10^{-4}$.}
    \label{fig:completeAS99}
\end{figure}

In this section, we present and compare the GBFs, spectra, Page evaporation factors, and computational times obtained using the two methods described in Sec. \ref{NumericalMethods}, and applied to the metrics introduced in Sec. \ref{Metrics}.\\
In the case of the first metric macro-group, we consider a test bench composed of five different BH configurations: the Schwarzschild BH, and two configurations each of the Hayward and Asymptotically Safe BHs, characterized by the values of the regularizing parameters $\ell=0.9 \times\sqrt{16/27}M$, $\ell=0.99 \times \sqrt{16/27}M$, $q=\frac{1.37}{0.9 \times M^2}$, and $q=\frac{1.37}{0.99 \times M^2}$. For brevity, we henceforth denote these configurations as Sch, H09, H99, AS09, and AS99, respectively.\\
For each configuration, we consider perturbations with spin $s=0, 1/2, 1$ and display the first three GBFs of each field, computed using both methods. Moreover, to better quantify the discrepancies between the two approaches, we consider the following estimator:
\begin{equation}\label{DeltaGamma}
\Delta(\Gamma)=\frac{|\Gamma_{s,l}^{\rm WKB}-\Gamma_{s,l}^{\rm GH}|}{\Gamma_{s,l}^{\rm GH}}.
\end{equation}
where $GH$ and $WKB$ denote quantities computed using GrayHawk and the WKB code, respectively.
By construction, $\Delta(\Gamma)$ is highly sensitive to numerical noise at low frequencies, since the GBFs tend to zero as $\omega M \rightarrow 0$. Consequently, in the low-frequency regime, $\Delta(\Gamma)$ amplifies numerical noise rather than providing an appropriate measure of the discrepancy between the two methods. We therefore introduce a cut-off when the GBF values computed with GrayHawk fall below $10^{-4}$. This threshold is one order of magnitude higher than the accuracy of the GrayHawk computation in the high-frequency regime, where the numerical accuracy is at its lowest.\\
\noindent Additionally, we consider two quantities derived directly from the GBFs: the primary emitted spectra, described by Eq. \ref{ParticleSpectrum}, and the Page evaporation factors, described by Eq. \ref{f_s}.\\
\noindent We organize these results into five figures and one table. Figures \eqref{fig:complete}--\eqref{fig:completeAS99} contain four plots each: a comparison between the two methods for the first three GBFs of each field, together with the corresponding $\Delta(\Gamma)$, and the primary emitted spectra computed for a BH with mass $M=10^{13}$ kg. Table 1 reports the values of the functions $f_s$ computed using the two methods, denoted by $f_s^{\rm GH}$ and $f_s^{\rm WKB}$, respectively, together with a percentage estimate of their discrepancy,
\begin{equation}
\Delta_s = \frac{|f_s^{\rm WKB}-f_s^{\rm GH}|}{f_s^{\rm GH}}\times100.
\end{equation}\\
\noindent We first note several features shared by all BH configurations and fields considered. The WKB approximation exhibits excellent agreement with the numerical computation in two independent regimes: at high frequencies and for large $l$, as expected. In fact, this is in accordance with what found in \cite{Lutfuoglu:2026vcr} in the context of Loop Quantum Gravity for a Holonomy-Corrected Schwarzschild BH \cite{Alonso-Bardaji:2022ear}, in \cite{Skvortsova:2026mwz} for a Dymnikova BH \cite{Dymnikova:1992ux}, and in \cite{Tan:2026vif} in the context of Lovelock gravity \cite{Lovelock:1971yv}.  The only appreciable discrepancies in the plots that show the functional GBFs form (in blue line results
computed with GrayHawk and in red those obtained using the WKB code), are for the $l=0$ mode, while those for the $l=1/2$ mode are barely appreciable. For all other cases the two methods visually overlap to such an extent that their differences can only be identified through $\Delta(\Gamma)$. Therefore, we now focus on the plots displaying $\Delta(\Gamma)$ for the first three GBFs of each field, in red, blue, and green, respectively. Such plots are characterized by a cut-off in the low energies regime when the GBFs values are comparable with the conservative value for the accuracy of GrayHawk: $10^{-4}$.
Regardless of the mode, for energies $\omega M \gtrsim 0.5$, the estimator satisfies $\Delta(\Gamma)\lesssim10^{-4}$ and is comparable to the numerical noise of the computation at high energies, confirming the excellent agreement between the two results. Conversely, in the low-energy region, the discrepancies become non-negligible even before reaching the cut-off region. The energy of the cut-off increases with increasing $l$ in the field decomposition. In all cases, the cut-off consistently depends only on the $l$-mode and occurs at approximately $\omega M \sim 0, 0.05, 0.1, 0.2, 0.3, 0.4, 0.5$ for $l=0,1/2,1,3/2,2,5/2,3$, respectively. This result is expected, since, to first approximation, the non-vanishing contribution of the GBFs shifts towards higher frequencies for increasing $l$-modes, independently of the BH solution.\\
Lower $l$-modes converge with high accuracy at lower values of $\omega M$. At low energies, the green curves lie above the blue ones, which in turn lie above the red ones, before converging towards values of approximately $10^{-4}$ when increasing $\omega$. At first glance, this behavior may appear to contradict the eikonal limit. However, this is not the case, since the most relevant contribution to each GBF is contained in its transient region, which is shifted towards larger frequencies for increasing $l$, while the asymptotic values are always 0 and 1. By rescaling the plots according to the central value and dispersion of each GBF functional profile, one clearly recovers the increased accuracy of the WKB approximation for higher $l$-modes. Thus, the relevant comparison is not to be taken at fixed frequency, but relative to the inflexion point of the corresponding $l$-mode.\\
\noindent We now focus on the plots displaying the spectra (in blue for the results computed using GrayHawk and in red for those obtained with the WKB code, with dotted, dashed, and solid lines corresponding to scalar, fermionic, and vector fields, respectively). Since the WKB approximation performs poorly at low energies, the spectra computed using this approximation deviate from the fully numerical results in this regime. More specifically, three distinct behaviors, common to all fields, can be identified. First, at high energies, the two spectra coincide and exhibit excellent agreement. At intermediate energies, the WKB spectra underestimate the particle flux. This underestimation appears to be alleviated by fermionic statistics. Finally, in the low-energy portion of the spectrum, the WKB method fails to reproduce the decreasing behavior of the spectral tails, thereby overestimating the spectral intensity.\\
Interestingly, in the nearly extremal cases, H99 and AS99, the WKB approximation does not allow the peak of the primary spectrum to be clearly identified. In the other cases, it remains possible to estimate the peak position, albeit with some uncertainty, by considering the local minimum between the high- and intermediate-energy regions. This behavior can be understood as a consequence of the interplay between the Hawking temperature and the accuracy of the WKB approximation. As extremality is approached, the Hawking temperature decreases, shifting the thermal distribution, and therefore the primary spectrum, towards lower frequencies. This is precisely the region where the WKB approximation exhibits its largest deviations from the fully numerical computed GBFs. As a consequence, in the nearly extremal configurations, the spectral functional shape is increasingly determined by a frequency range in which the WKB approximation is less accurate, making the peak position difficult to identify reliably.\\
\noindent A similar behavior is observed for the Page evaporation factors reported in Table \ref{tab:page_factors}.

\begin{table}[H]
\centering
\begin{tabular}{llccc} 
\hline
Metric & $s$ & $f_s^{\rm GH}$ & $f_s^{\rm WKB}$ & $\Delta_s$ [\%] \\
\hline
Sch
& $0$   & $7.4346\times10^{-5}$ & $6.4900\times10^{-5}$ & $12.70$ \\
& $1/2$ & $4.0916\times10^{-5}$ & $3.8658\times10^{-5}$ & $5.52$ \\
& $1$   & $1.6819\times10^{-5}$ & $1.6799\times10^{-5}$ & $0.12$ \\
\hline
H09
& $0$   & $6.5525\times10^{-6}$ & $3.6182\times10^{-6}$ & $44.78$ \\
& $1/2$ & $1.1781\times10^{-6}$ & $1.0988\times10^{-6}$ & $6.73$ \\
& $1$   & $1.2104\times10^{-7}$ & $1.2486\times10^{-7}$ & $3.16$ \\
\hline
H99
& $0$   & $8.2039\times10^{-8}$ & $4.3251\times10^{-8}$ & $47.28$ \\
& $1/2$ & $1.2979\times10^{-9}$ & $8.1495\times10^{-9}$ & $527.92$ \\
& $1$   & $1.0538\times10^{-11}$ & $7.7557\times10^{-10}$ & $7259.88$ \\
\hline
AS09
& $0$   & $2.0584\times10^{-6}$ & $8.7759\times10^{-7}$ & $57.36$ \\
& $1/2$ & $2.0481\times10^{-7}$ & $1.8359\times10^{-7}$ & $10.36$ \\
& $1$   & $1.0428\times10^{-8}$ & $1.2661\times10^{-8}$ & $21.41$ \\
\hline
AS99
& $0$   & $1.2028\times10^{-8}$ & $1.8154\times10^{-8}$ & $50.92$ \\
& $1/2$ & $8.4927\times10^{-11}$ & $7.9241\times10^{-10}$ & $833.05$ \\
& $1$   & $2.8059\times10^{-13}$ & $2.1146\times10^{-10}$ & $75262.37$ \\
\hline
\end{tabular}
\vspace{0.7cm}
\caption{Page evaporation factors for different spin fields ($s=0,1/2,1$), obtained with GrayHawk and the WKB code. In the last column, the relative difference between the two methods is expressed as a percentage, $\Delta_s = \frac{|f_s^{\rm WKB}-f_s^{\rm GH}|}{f_s^{\rm GH}}\times100$.}
\label{tab:page_factors}
\end{table}
\noindent In the Schwarzschild case, it is evident that, as the spin of the field increases, the WKB-computed factors converge towards the fully numerical results. This can be understood from the fact that the computation of $f_s$ is dominated by the lowest $l$-mode, whose value increases with the spin of the field, thereby making the WKB approximation increasingly accurate.\\
Similarly, H9 and AS9 display comparable behavior, with the exception of the vector field in AS9, which deviates from the observed convergence trend. The same deviation is found for H99 and AS99. In these cases, higher-spin fields are characterized by a worsening agreement between the WKB-computed Page evaporation factors and the fully numerical results. We hypothesize that this trend is related to the more complex potential characterizing the nearly extremal solutions, which cannot be accurately represented within the WKB approximation.\\
\noindent Finally, we comment on the computational time required for a single GBF calculation with the two codes. {We show the computational time for the GBFs for all the metric configurations considered through the paper, with $s=l$, in Table \ref{computational_times2}. } The WKB code is nearly instantaneous and, when executed, provides a GBF in a fraction of a second. Conversely, the GrayHawk runtime depends on the complexity of the metric and can range from a few tens of seconds to several minutes.

\begin{table}[h]
\centering
\begin{tabular}{llcc}
\hline
Metric & $s$ & GrayHawk & WKB \\
\hline
Sch  & $0$   & 30.279  & 0.376 \\
     & $1/2$ & 38.234  & 0.336 \\
     & $1$   & 31.618  & 0.261 \\
\hline
H09  & $0$   & 107.812 & 0.392 \\
     & $1/2$ & 131.180 & 0.310 \\
     & $1$   & 109.986 & 0.296 \\
\hline
H99  & $0$   & 99.611  & 0.391 \\
     & $1/2$ & 127.941 & 0.331 \\
     & $1$   & 115.239 & 0.315 \\
\hline
AS09 & $0$   & 40.514  & 0.401 \\
     & $1/2$ & 48.614  & 0.322 \\
     & $1$   & 38.696  & 0.301 \\
\hline
AS99 & $0$   & 38.201  & 0.407 \\
     & $1/2$ & 50.435  & 0.317 \\
     & $1$   & 41.029  & 0.310 \\
\hline
\end{tabular}
\vspace{0.7cm}
\caption{Computational times, in seconds, required to obtain the $l=s$ GBF using the GrayHawk and sixth-order WKB methods.}
\label{computational_times2}
\end{table}

\begin{figure}[H]
    \centering

    \begin{subfigure}[b]{0.55\textwidth}
        \centering
        \includegraphics[width=0.9\textwidth, height=5.5cm]{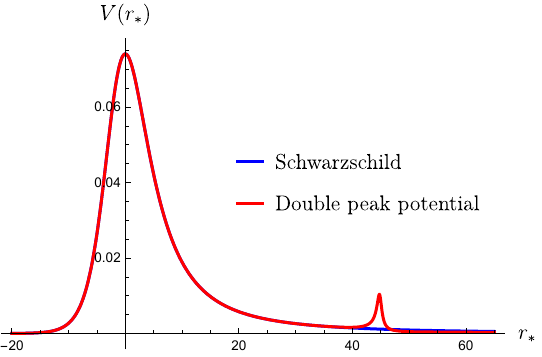}
        \caption{Potential.}
        \label{Doublepotential}
    \end{subfigure}
    \hfill
    \begin{subfigure}[b]{0.44\textwidth}
        \centering
        \includegraphics[width=0.9\textwidth, height=7cm]{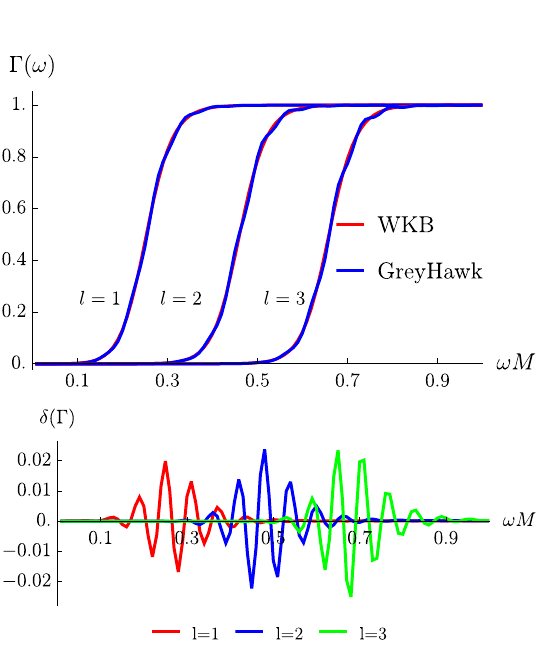}
        \caption{Photon.}
        \label{photondouble}
    \end{subfigure}

    \caption{This figure shows the results for the geometry given by (\ref{eq:multi_peak_metric}). The red curve of panel(a) is the effective potential given by the metric function \eqref{eq:multi_peak_potential} for $s=l=1$ and the bllue curve is the effective potential for the Schwarzschild case. Panel(b) show the GBFs and corresponding $\delta(\Gamma)$, as defined in \eqref{deltagamma} for massless fields of spin $s=1$ and angular number $l=1,2,3$. In blue are reported the results computed using GrayHawk, while the 6th-order WKB method is reported in red. The $\delta(\Gamma)$ plots report, in red, blue, and green, the errors of modes $l=1,2,3$, respectivly.}
    \label{fig:DoublePeak}
\end{figure}

\newpage
\noindent
We now turn our attention to the second macro-group of metric configurations. In this case, we consider electromagnetic perturbations ($s=1$) and compute the GBFs for the multipole modes $l=1,2,3$, adopting the metric described by Eq. \eqref{eq:multi_peak_metric} and setting $\eta=0.01$, $\sigma=5$, and $r_e=50$. The left panel of Fig. \ref{fig:DoublePeak} displays the corresponding effective potential for $s=l=1$ (red curve), compared with the Schwarzschild case (blue curve). As expected from the construction of the metric function in Eq. \eqref{eq:multi_peak_metric}, the effective potential exhibits a double-peaked structure in tortoise coordinates. In particular, the primary peak, which is also present in the Schwarzschild configuration, is located approximately at $r_*=0$, while a smaller secondary peak, absent in the Schwarzschild case, emerges around $r_*=r_e$.\\
The upper-right panel of Fig. \ref{fig:DoublePeak} shows the GBFs for massless electromagnetic perturbations with $l=1,2,3$, computed using the WKB code (red curves) and the fully numerical code GrayHawk (blue curves). The two approaches yield GBFs that agree to a good approximation at the qualitative level. However, a closer inspection reveals an oscillatory behaviour in the numerically computed GBFs, which is not reproduced by the WKB results. This discrepancy can be attributed to the limitations of the standard WKB approximation when applied to multi-peaked effective potentials. Indeed, the conventional WKB prescription estimates the GBFs by expanding the potential around its maximum and matching the corresponding solutions across the relevant turning points. Such a procedure does not fully account for the successive scattering processes induced by the presence of multiple barriers, where partial reflection and transmission lead to repeated interactions between the different potential peaks. Consequently, an approach based solely on the global maximum cannot capture the interference effects associated with these multiple scattering events. In the present configuration, this limitation results in WKB GBFs that remain very close to those obtained in the Schwarzschild spacetime. By contrast, the fully numerical calculations performed with GrayHawk capture the oscillatory features arising from repeated reflections within the partially confining region between the two barriers, which are responsible for the emergence of echo-like contributions (it must be noticed that the echo-like contribution is shared with every compact object whose potential is characterized by 2 peaks \cite{Bueno:2017hyj,Cardoso:2016oxy,Churilova:2021tgn,Maggio:2019zyv,Maggio:2020jml,Aneesh:2018hlp,Rosato:2025byu}).\\
To quantify the limitations of the WKB approximation in reproducing these features, we introduce the relative discrepancy $\delta(\Gamma)$
\begin{equation} \label{deltagamma}
    \delta(\Gamma)=\Gamma^{WKB}_{s,l}-\Gamma^{GH}_{s,l}
\end{equation}
and display its behaviour for the three multipole modes considered in the lower-right panel of Fig. \ref{fig:DoublePeak}. For the specific configuration under consideration, the discrepancy reaches values of approximately $2-3\%$ relative to the maximum value of the GBFs. We stress, however, that this result is strongly dependent on the adopted metric parameters and, in particular, on the height, width, and relative separation of the secondary peak. Accordingly, $\delta(\Gamma)$ is not expected to remain uniformly small across the entire parameter space. Indeed, we verified that it can reach values of order unity for configurations in which the primary and secondary peaks have comparable heights and widths and are sufficiently close to one another.

\section{Conclusion}
\label{Conclusion}
In this work, we have presented a systematic comparison between two independent approaches for the computation of Gray-Body Factors (GBFs): the fully numerical GrayHawk code and a code based on the sixth-order WKB approximation. We first have considered three different Black Hole (BH) geometries, namely the Schwarzschild, Hayward, and Asymptotically Safe solutions, and investigated five representative configurations spanning from non-extremal to nearly extremal regimes. For each geometry, we have considered massless scalar, fermionic, and vector perturbations and compared the resulting GBFs, primary emission spectra, and Page evaporation factors.\\
The comparison of the GBFs confirms the expected behavior of the WKB approximation. At high frequencies and, more generally, in the regime of large $l$, the WKB results are in excellent agreement with the fully numerical computation. In particular, for $\omega M\gtrsim0.5$, the relative discrepancy is typically of the order of $10^{-4}$, comparable with the numerical accuracy of GrayHawk in this regime. The main deviations occur at low frequencies and are most pronounced for the lowest multipolar modes. The characteristic frequency below which the two methods begin to differ increases with $l$, reflecting the shift of the peak of the effective potential and, as a consequence, the relevant contribution of the GBFs towards higher frequencies for higher multipoles.\\
The comparison of the emission spectra further highlights the different regimes of validity of the two approaches. At high energies, the spectra obtained with the two methods are essentially indistinguishable. At intermediate energies, the WKB approximation tends to underestimate the particle flux, while at low energies it fails to reproduce the decreasing behavior of the spectral tails and consequently overestimates the spectral intensity. These discrepancies become particularly significant when the quantities derived from the GBFs are integrated over the full frequency range, as in the computation of the Page evaporation factors.\\
A particularly relevant outcome of our analysis is the emergence of two qualitatively different BH regimes. The first corresponds to the non-near-extremal configurations, represented by Schwarzschild, H09, and AS09. In this regime, the WKB approximation generally provides a reliable description of the GBFs and of the derived observables, especially for higher-spin fields and higher multipoles. For the Schwarzschild solution, for example, the agreement between the WKB and fully numerical Page factors improves systematically with increasing field spin, since the lowest contributing multipoles become increasingly well described within the WKB regime. H09 and AS09 exhibit a broadly similar behavior, although some deviations are already visible for the vector field in AS09.\\
The second regime is represented by the nearly extremal configurations H99 and AS99. Here, the discrepancies between the two approaches become substantially larger, particularly for higher-spin fields. The WKB approximation can fail to reproduce even the position of the peak of the primary emission spectrum and yields increasingly large deviations in the Page evaporation factors. This behavior can instead be understood in terms of the interplay between the Hawking temperature and the frequency dependence of the WKB accuracy. As extremality is approached, the Hawking temperature decreases, shifting the primary spectrum towards lower frequencies, where the WKB approximation exhibits its largest deviations from the fully numerical GBFs. As a result, the spectral peak becomes increasingly sensitive to the low-frequency inaccuracies of the WKB approximation, leading to larger discrepancies in the derived Page factors. The dramatic increase in the discrepancies of the Page factors for higher-spin fields in H99 and AS99 provides further evidence that the near-extremal regime constitutes a distinct domain in which the fully numerical treatment is essential. A similar conclusion has been drawn in \cite{Lutfuoglu:2026zxj} in the case of a Simpson–Visser BH \cite{Simpson:2018tsi}.
If an increase in the regularizing parameter lowers the BH temperature, the distinction between non-near-extremal and near-extremal BHs is important when assessing the applicability of the WKB approximation. In the non-near-extremal regime, the dominant contribution to the emission spectrum lies predominantly in a frequency range where the WKB approximation accurately reproduces the GBFs. As extremality is approached, the deviations of the WKB approach grow larger. This effect is particularly important for integrated quantities such as the Page evaporation factors. We underline that this reasoning does not apply to BH configurations whose surface gravity, and therefore temperature, increase as the regularizing parameter increases \cite{Zhang:2024ney}, which, close to extremality, should exhibit a different behavior.\\
Interestingly, the two approaches differ substantially in computational cost. The WKB code provides a GBF essentially instantaneously, requiring only a fraction of a second for a single computation, whereas GrayHawk requires from a few tens of seconds to several minutes, depending on the complexity of the metric. This difference makes the WKB approach particularly attractive when a large number of GBFs must be computed rapidly, especially in the high-energy or high-$l$ regime where its accuracy is highest. Conversely, although GrayHawk is computationally more demanding, its direct numerical treatment preserves the accuracy of the GBFs in the low-energy regime and therefore provides a more reliable description of the low-energy emission spectrum and of observables sensitive to this region.\\
\noindent We have also investigated the impact of multi-peaked effective potentials on the scattering properties of electromagnetic perturbations, focusing on a metric configuration characterized by the presence of two distinct potential maxima. By computing the corresponding GBFs for different multipole modes, we have shown that the secondary peak introduces non-trivial modifications to the scattering process. In particular, the fully numerical results exhibit an oscillatory structure that can be interpreted as the outcome of successive reflections and transmissions between the two potential barriers. These multiple-scattering effects, associated with the partial confinement of the perturbations in the region separating the peaks, give rise to echo-like features in the GBFs. We have further demonstrated that the standard WKB approximation, based on an expansion of the effective potential around its dominant maximum, is unable to fully reproduce these oscillatory patterns. While the WKB and numerical results remain in good agreement for the specific configuration considered, discrepancies can become substantially larger depending on the relative height, width, and separation of the potential peaks. In particular, configurations featuring two comparable and sufficiently close maxima can lead to significant deviations, highlighting the limitations of conventional WKB-based estimates in the presence of complex effective-potential structures. These results emphasize the importance of employing fully numerical approaches, or more general semi-analytical methods accounting for multiple scattering, when investigating wave propagation in spacetimes characterized by non-trivial effective potentials.\\
Overall, our results indicate that the two methods should be regarded as complementary rather than competing approaches. The WKB approximation offers a fast and accurate tool for high-energy calculations, while GrayHawk provides the more robust approach when low-frequency behavior, or high-precision integrated observables are of interest. The combination of the two methods therefore offers an efficient strategy for exploring Hawking emission across a broad range of BH geometries and physical regimes.

\section*{Acknowledgments}

We would like to acknowledge R. Konoplya and A. Zhidenko for useful discussions and suggestions.

\clearpage

\bibliographystyle{JHEP}
\bibliography{biblio}

\end{document}